\documentclass[epj]{svjour}
\usepackage{graphics}
\usepackage{mathrsfs}
\usepackage{amsmath,amssymb,graphics,epsfig,subfigure}
\usepackage{color}
\begin{document}

\title{On the correspondence between energy conservation and energy-momentum tensor conservation in cosmology}

\author{Hao Yu\inst{1} \and Bao-Min Gu\inst{2,3} \and Zhi Luo\inst{1} \and Jin Li\inst{1}}
%

\institute{College of Physics, Chongqing University, Chongqing 401331, China \and
Department of Physics, Nanchang University, Nanchang 330031, China \and
Center for Relativistic Astrophysics and High Energy Physics, Nanchang University, Nanchang, 330031, China}         

\date{Received: date / Revised version: date}
%
\abstract{
The correspondence between the thermodynamic energy equation satisfied by a closed co-moving volume and the conservation equation satisfied by the energy-momentum tensor of the matter inside the co-moving volume is extended to a more general system with an arbitrary cosmological horizon and a heat source. The energy of the system consisting of a cosmological horizon and its internal matter could be conserved by defining a surface energy on the horizons. Therefore, energy conservation and energy-momentum tensor conservation can always be consistent for such a system. On the other hand, from the perspective of classical thermodynamics, one can define an effective pressure at the cosmological horizon to guarantee that the thermodynamic energy equation inside the horizon is consistent with the energy-momentum tensor conservation equation of the matter inside the horizon. These systems can satisfy the generalized second law of thermodynamics under appropriate conditions. The definitions of the surface energy and the effective pressure are extended to the gravity theory with non-minimal coupling between geometry and matter, in which geometry could be regarded as a heat source.
}
\maketitle
\thanks{\emph{Present address:} cqujinli1983@cqu.edu.cn}

\section{Introduction}

Classical thermodynamics is of great significance to the development of physics, whose validity has been verified in different fields, including astrophysics and cosmology. In the 1970s, physicists represented by Bekenstein and Hawking, in the quest for describing black holes, discovered that the evolution equations of certain gravitational phenomena are similar to the laws and equations of thermodynamics~\cite{Bekenstein:1972tm,Bekenstein:1973ur,Hawking:1974rv,Hawking:1975vcx,Bardeen:1973gs}, which provides new inspirations for revealing the deep correlation between general relativity (GR) and quantum field theory (see Refs.~\cite{Wald:1999vt,Carlip:2014pma} for comprehensive reviews of black hole thermodynamics). Then Gibbons and Hawking developed the research in cosmology and calculated the temperature and entropy of cosmological event horizons~\cite{Gibbons:1977mu}. In the 1990s, 't Hooft and Susskind attempted to study holographic principle with the (second) laws of black hole thermodynamics, and advocated that quantum gravity theories should be holographic~\cite{tHooft:1993dmi,Susskind:1994vu,Aharony:1999ti}. Against the backdrop of the research, the connection between thermodynamic laws and the properties of gravitational systems has attracted more and more attention~\cite{Padmanabhan:2002sha,Jacobson:1995ab,Eling:2006aw,Hayward:1997jp,Mukohyama:1999sp,Cai:2005ra,Cai:2006rs,Akbar:2006kj,Padmanabhan:2003gd,Padmanabhan:2009vy,Gong:2007md}.

In fact, the research on thermodynamics in gravitational systems could be traced back to the first half of the last century~\cite{Tolman11}. Since in the 1920s, astronomical observations showed that the universe is expanding, naturally one can study thermodynamics in the context of cosmology by replacing directly the expanding volume in classical thermodynamics with the co-moving volume. In this case, the thermodynamic energy equation corresponding to the first law of thermodynamics for a closed system (such as the co-moving volume) is consistent with the energy-momentum tensor conservation equation of the matter in cosmology. In the 1980s, Prigogine et al. began to systematically study particle production and the thermodynamic energy equation in the co-moving volume. By comparing the thermodynamic energy equation of an open expanding system with the cosmological evolution equation with an irreversible matter creation process, they summarized the thermodynamic and cosmological properties of particle production~\cite{Prigoginex1,Prigoginex2}. About ten years ago, Harko et al., based on previous research, compared the thermodynamic energy equation of an open co-moving volume with the evolution equation of energy-momentum tensor in the gravity theory with non-minimal coupling between geometry and matter~\cite{Harko:2008qz,Bertolami:2008zh,Harko:2011kv}. They found that the non-conservation of the energy-momentum tensor of the matter can be explained as irreversible particle production in an open system~\cite{Harko:2014pqa,Harko:2015pma}. In general, the core of these studies is the correspondence between (thermodynamic) energy conservation and energy-momentum tensor conservation. Due to the Bianchi identities, energy-momentum tensor conservation always holds in GR. Energy conservation is a pillar in classical physics, but it encounters troubles in GR and quantum mechanics. Is energy conserved in GR? The answer to this question is controversial. The definition of localized energy density in GR is still a subtle problem~\cite{Szabados:2009eka}, which has not been completely solved even though GR has been proposed for over a hundred years. In the early research on GR, Einstein, Landau, Lifshitz and others tried to define the energy-momentum tensor of the gravitational field so that the total energy of matter fields and the gravitational field could be conserved. However, the energy-momentum tensors of the gravitational field constructed in this way are all pseudotensors which are coordinate dependent (see Refs.~\cite{Szabados:2009eka,Coller:1958tx,Arnowitt:1959ah,Arnowitt:1961zz,Maluf:1995re} and references therein for more detail). Actually, the equivalence principle implies that we can transform away gravity when we move from one frame to another, and hence there cannot be such an energy-momentum tensor of the gravitational field which is independent of coordinate. Given that, the current consensus among physicists is that the law of energy conservation in GR is not valid.

In this paper, we continue to discuss energy conservation and energy-momentum tensor conservation in cosmology based on the laws of thermodynamics. The volume of the system will be extended from the conventional co-moving volume to the volume enclosed by an arbitrary cosmological horizon~\cite{Rindler:1956yx,Rindlerb}. By defining a surface energy on the cosmological horizon, the energy conservation of the system can be ensured, so that the energy conservation equation and the energy-momentum tensor conservation equation can be consistent. In addition, by defining an effective pressure at the cosmological horizon, we may also ensure that the thermodynamic energy equation of the matter inside the horizon is consistent with the corresponding energy-momentum tensor conservation equation. Both of these correspondences can be regarded as the correspondence between energy conservation and energy-momentum tensor conservation. For these cosmological horizons, we verify whether the system satisfies the generalized second law of thermodynamics~\cite{Bekenstein:1974ax,Davies:1987ti,Wu:2007se,Wu:2008ir}. For the gravity theory with non-minimal coupling between geometry and matter (non-minimal coupling gravity), the evolution equation of the energy-momentum tensor of the matter inside the cosmological horizon could correspond to the thermodynamic energy equation of an open system with a heat source~\cite{Prigoginex1,Prigoginex2}.

Usually the energy of the (apparent) horizon refers to the energy exchanged between the inside and outside of the horizon during a certain period of time~\cite{Cai:2005ra,Cai:2006rs,Akbar:2006kj,Frolov:2002va}. Here, the surface energy on the horizon that we define could be on account of the tension on the horizon or the intrinsic property of the horizon. Moreover, when there exists particle production in the universe, we can consider the impact of particle production on the evolution of the universe as an extra pressure~\cite{Ford:1986sy,Traschen:1990sw,Abramo:1996ip,Zimdahl:1999tn}, so the effective pressure of the universe is equal to the pressure of the matter plus the extra pressure related to particle production. The effective pressure at the horizon that we define may also be caused by an unknown cosmological process or the intrinsic property of gravity.

Our research motivation is twofold. On the one hand, although the co-moving volume is the system most similar to the classical thermodynamic closed system (in which the particle number is conserved in the absence of particle production and annihilation), its boundary, which may be called co-moving horizon, has no specific physical meaning in cosmology. However, other cosmological horizons have clear physical meanings. For example, the apparent horizon is closely related to holographic principle and the thermodynamic behavior of the universe~\cite{Akbar:2006kj,Bousso:2002ju,Bak:1999hd,Faraoni:2011hf}, and the cosmic event horizon is the largest co-moving distance that a photon emitted at present can propagate in the future~\cite{Giovannini:2008zzb,Davis:2003ad}. Therefore, studying energy conservation and energy-momentum tensor conservation for different cosmological horizons may reveal more characteristics of various horizons. On the other hand, the introduction of the surface energy on the horizon can solve the energy conservation problem of the universe from a phenomenological perspective, so it may also provide an alternative to define the local energy density of the gravitational field in GR. The definition of the effective pressure at the horizon, similar to particle production pressure, may explain the expansion of the universe even without dark energy~\cite{Ford:1986sy,Traschen:1990sw,Abramo:1996ip,Zimdahl:1999tn}.

The paper is organized as follows: Sec.~\ref{sec2} is devoted to introducing energy conservation in cosmology and analyzing why it cannot be implemented by bringing in the non-minimal coupling between matter fields or defining the local energy density of the gravitational field. In Sec.~\ref{sec3}, we study how to implement energy conservation in cosmology by defining a surface energy on a cosmological horizon or an effective pressure at the horizon. With these definitions, the energy conservation equation (or the thermodynamic energy equation) of the system is consistent with the energy-momentum tensor conservation equation of the matter. In Sec.~\ref{sec4}, we calculate and analyze whether the total entropy of the system satisfies the generalized second law of thermodynamics. In Sec.~\ref{sec5}, we extend these studies to the gravity theory with non-minimal coupling between geometry and matter. The discussions and conclusions are rendered in Sec~\ref{sec6}.

\section{Energy conservation in cosmology}
\label{sec2}
In cosmology, whether the thermodynamic energy equation can correspond to the energy-momentum tensor conservation equation, is dependent on whether the energy of the system is conserved. Only when the energy of the universe is conserved, one can derive the energy-momentum tensor conservation equation ($\nabla_\nu T^{\nu\mu}=0$) from the thermodynamic energy conservation ($\text{d}E=-p\,\text{d}V$)~\cite{Tolman11}. The approach here is to default that the selected system (i.e., the co-moving volume) can do work on its surroundings, and the effective pressure at the boundary of the system is equal to the pressure of the internal substance. Such correspondence can be extended to the open system with a heat source or particle production in the non-minimal coupling gravity, i.e., $\nabla_\nu T^{\nu\mu}=\hat X^\mu$ corresponds to $\text{d}E=\text{d}Q-p\,\text{d}V$~\cite{Prigoginex1,Prigoginex2}. In this section, we briefly review energy conservation in cosmology and then analyze, from a phenomenological perspective, why energy conservation in the context of cosmology cannot be preserved by introducing the non-minimal coupling between matter fields or defining the energy density of the gravitational field~\cite{Moller1,Moller2}.

The most intuitive phenomenon about energy non-conservation in the universe is the red-shift of photons. According to the observations of the cosmic microwave background (CMB)~\cite{Penzias:1965wn,Fixsen:2009ug}, the CMB photons are red-shifted as the universe expands. Since the CMB has no interaction with other matter fields after photon decoupling (i.e., the CMB photons are free to travel), it is confusing where the energy of the CMB photons flows after photon decoupling.

First, we can interpret the energy loss of photons as the work done by the universe on its surroundings ($\text{d}E=-p\,\text{d}V$). In this case, we can ensure energy conservation (for the system and outside), while $\text{d}E=-p\,\text{d}V$ and $\nabla_\nu T_p^{\nu\mu}=0$ (where $T_p^{\nu\mu}$ is the energy-momentum tensor of the CMB photons) are consistent. However, for the photons traveling completely freely, the expansion of the universe is more similar to the free expansion in classical thermodynamics, so the energy of photons should not be lost ($\text{d}E=0$)\footnote{Physicists hold the opinion that the expansion of the universe is the intrinsic property of space-time rather than the widening of the boundary. The assumption that the boundary of the co-moving volume could do work on its surroundings is just a special way to study the thermodynamics of the universe~\cite{Prigoginex1,Prigoginex2,Harko:2014pqa,Harko:2015pma}. Regarding the intrinsic property of space-time and whether the expansion of the universe should be  completely free, we will not discuss these problems further here.}. Note that the correspondence between $\text{d}E=\text{d}Q-p\,\text{d}V$ and $\nabla_\nu T^{\nu\mu}=\hat X^\mu$ in Refs.~\cite{Prigoginex1,Prigoginex2,Harko:2014pqa,Harko:2015pma} is a phenomenological result. So far we do not figure out if there are deeper reasons for the consistency. Therefore, at least from the perspective of the CMB photons, interpreting their energy loss by doing work requires us to discard the standpoint that the universe is expanding freely. In our following research, we will also use the concept of work to explain the energy transformation of the universe.

Moreover, if we can well define the energy density of the gravitational field, it is possible to suppose that there exists energy exchange between the gravitational field and the CMB photons. Therefore the energy of photons may flow into the gravitational field. However, such a definition has been proven to be impracticable or defective in GR, and so it is not likely to be feasible in cosmology. Noether's theorem tells us that energy conservation is the result of time translation symmetry. Since there is no time translation symmetry in the expanding universe, there is no energy conservation in cosmology according to classical Noether's theorem. From the perspective of GR, the ``Noether current'' related to the gravitational field (the energy-momentum tensor of the gravitational field) is a pseudo-tensor $\tilde T^{\mu\nu}$ with two indices. Although we have $\nabla_\nu \tilde T^{\mu\nu}=0$, when we try to use Gauss theorem to convert this local conservation law into a global conservation law (energy conservation), the presence of the Christoffel symbols prevent us from doing that in a valid way for all frames ($\tilde T^{\mu\nu}$ must be dependent on the coordinate system)~\cite{Szabados:2009eka,Coller:1958tx,Arnowitt:1959ah,Arnowitt:1961zz,Maluf:1995re}. Such a property is essentially different from the ``Noether current'' corresponding to the electromagnetic field. Therefore, for most cosmologists the non-conservation of energy in the universe is not ``riddle'', because it is still perfectly true that energy-momentum tensor is always conserved locally and covariantly. Regarding the issue of energy conservation in cosmology and GR, one can abandon it and think that energy is indeed not conserved, but if one insists on energy conservation from a phenomenological perspective, one might see something inspiring especially in cosmology. Next we analyze how to implement energy conservation in cosmology from the perspective of phenomenology.

Since energy non-conservation in cosmology is mainly manifested in the red-shift of the CMB photons, we start from the CMB photons and analyze whether a certain component of the universe can be coupled to photons through an unknown form, so as to explain the energy loss of photons on a phenomenological level. For our homogeneous and isotropic universe, the metric can be given as
\begin{eqnarray}
\text{d}s^2=-c^2\,\text{d}t^2+a^2(t)(\text{d}x^2+\text{d}y^2+\text{d}z^2),
\end{eqnarray}
where $a(t)$ is the scale factor and we have assumed that the curvature of the universe is zero. Therefore, Friedmann equations read
\begin{eqnarray}
3H^2=&&\!\!\kappa^2(\rho_d+\rho_r+\rho_{DM}
+\rho_{DE})\label{eq:FriedmannEq},\\
3H^2+2\dot{H}=&&\!\!-\kappa^2 (p_d+p_r+p_{DM}+p_{DE})\label{eq:PressureEq}\,,\,
\end{eqnarray}
where $\kappa^2\equiv {8\pi\, G}$, $H=\dot{a}/a$ is the Hubble rate, and the dot represents the derivative of time. The indexes $d, r, DM$, and $DE$ indicate dust, radiation (photons and neutrinos), (cold) dark matter, and dark energy, respectively. Their equations of state are given as $p_d=0$, $p_r=1/3\,\rho_r$, $p_{DM}=0$, and $p_{DE}=f(\rho_{DE})$. In the standard $\Lambda$CDM, $p_{DE}=-\rho_{DE}$.

Without loss of generality, we will study the system that is covered by a spherical shell with a time-evolving radius. The spherical shell could be referred to a cosmological horizon. Different evolution of the horizon radius represents different physical meaning. For example, if the horizon radius is proportional to the scale factor, it represents the boundary of the co-moving volume, which could be called as co-moving horizon. If the horizon is a surface that is the boundary between light rays that are directed outwards and moving outwards, and those directed outward but moving inward, it represents the apparent horizon~\cite{Wald:1991zz}. Here we define the energy inside a cosmological horizon as Misner-Sharp (MS) energy~\cite{Misner:1964je,Hayward:1993ph,Hayward:1994bu,Cai:2008gw} and suppose temporarily that the gravitational field has no energy, so the total energy of the universe inside the horizon is defined as
\begin{eqnarray}
\mathfrak{E_t}=(\rho_d+\rho_r+\rho_{DM}+\rho_{DE})V_h,
\end{eqnarray}
where $V_h$ is the volume inside the horizon. In the standard $\Lambda$CDM model, the energy-momentum tensor of photons after photon decoupling satisfies $\nabla_{\mu}T_p^{\mu\nu}=0$ and so the energy density of photons evolves as $\rho_{p}(a)=\rho_{p0}\,a^{-4}$, where $\rho_{p0}$ denotes the radiation density at present ($a=1$). Therefore, the MS energy corresponding to photons is
\begin{eqnarray}
\mathfrak{E}_p=\rho_p\,V_h=\rho_{p0}\,a^{-4}\,V_h.
\end{eqnarray}
For the co-moving volume ($V_h=\frac{4\pi}{3}a^3$) or the volume covered by the apparent horizon ($V_h=\frac{4\pi}{3}H^{-3}$ for vanishing curvature), $\mathfrak{E}_p$ is not a constant and so the energy of photons is non-conserved. Only when $V_h\sim a^{4}$, $\mathfrak{E}_p$ is conserved, but the photon number is not conserved. For a single photon, $\rho_{p}=\rho_{p0}\,a^{-4}$ means its wavelength will be elongated (shortened) in an expanding (collapsing) universe. Note that
$\rho_{p}=\rho_{p0}\,a^{-4}$ is robust since it is the direct consequence from the CMB observation. Therefore, the red-shift of a single photon in the universe is deemed to be a manifestation of the energy non-conservation in GR. If one supposes that energy non-conservation is not a sacred theorem (for example, energy conservation in quantum mechanics is not a rule that must be guarded), the red-shift of a single photon is no longer a problem. But it is still difficult to stop asking where the energy has gone.

Let us inspect whether other components of the universe have similar energy non-conservation. For dust and (cold) dark matter, since their pressure is approximate to 0, in the $\Lambda$CDM model, their total energy density can be given as $\rho_{dDM}(a)=\rho_d+\rho_{DM}=\rho_{dDM0}\,a^{-3}$, where $\rho_{dDM0}$ is the total energy density of dust and (cold) dark matter at present. Therefore, the corresponding MS energy inside the horizon is
\begin{eqnarray}
\mathfrak{E}_{dDM}=\rho_{dDM}\,V_h=\rho_{dDM0}\,a^{-3}\,V_h.
\end{eqnarray}
For the co-moving volume ($V_h=\frac{4\pi}{3}a^3$), energy conservation can be satisfied and the total number of particles is also conserved. But for other volume (such as the apparent horizon), neither total energy nor the number of particles is conserved.

As for dark energy, assuming that the energy density is given as $\rho_{DE}(a)=\rho_{DE0}\,F(a)$, where $\rho_{DE0}$ denotes the current dark energy density and $F(a)$ is a function of the scale factor satisfying $F(1)=1$, the energy inside the horizon $V_h$ is give by
\begin{eqnarray}
\mathfrak{E}_{DE}=\rho_{DE}\,V_h=\rho_{DE0}\,F(a)\,V_h.
\end{eqnarray}
Obviously, the condition for energy conservation is $V_h=F^{-1}(a)$. But whether the number of particles is conserved is unknown.

Now let us check where the energy of photons may flow. The mainstream view is that it flows into the gravitational field. However, since the energy density of the gravitational field cannot be defined well, the explanation does not satisfy many people. Some researchers believe that it may flow into other matter fields, such as dark matter or dark energy, so as to ensure the total energy conservation of the universe. The current research on photons coupling to dark matter or dark energy~\cite{Ullio:2002pj,Chen:2003gz,Galli:2009zc,Wilkinson:2013kia,Boehm:2014vja,Kumar:2018yhh,Escudero:2018thh,Opher:2004vg,Opher:2005px} has seldom paid attention to the issue of the total energy of the universe. Next, we focus on the coupling between photons and other matter fields. Our choice of the horizon is still random. If the energy loss of photons can be explained by introducing the coupling, then the energy loss (or increase) of other components of the universe is no longer a problem.

Assume that photons are coupled to an unknown substance ($T_s^{\mu\nu}$) and other matter ($T_m^{\mu\nu}$) is temporarily ignored (i.e., $\nabla_{\mu}T_m^{\mu\nu}=0$ and $\rho_m\,V_h=\text{Const.}$). The equation of state of the (unknown) substance is $p_s=G(\rho_s)$, where $G(\rho_s)$ is a reasonable function of the energy density. Combining the contracted Bianchi identities (which results in $\nabla_{\mu}T_{total}^{\mu\nu}=0$) and $\nabla_{\mu}T_m^{\mu\nu}=0$, we have $\nabla_{\mu}(T_p^{\mu\nu}+T_s^{\mu\nu})=0$. Since the CMB obeys the law of black-body radiation, so $\nabla_{\mu}T_p^{\mu\nu}=0$ and $\nabla_{\mu}T_s^{\mu\nu}=0$. The corresponding energy densities are given as
\begin{eqnarray}
\rho_p=\rho_{p0}\,a^{-4},\\
-\frac13\int\frac{\text{d}\rho_s}{G(\rho_s)+\rho_s}=\ln a.\label{rhosa}
\end{eqnarray}
In order to ensure that the total energy of the universe is conserved within the horizon $V_h$, we need the energy densities of photons and the substance to satisfy the following relationship:
\begin{eqnarray}
(\rho_{p0}\,a^{-4}+\rho_s)V_h=\text{Const.},\label{eq210}
\end{eqnarray}
where $\rho_s$ is given by Eq.~(\ref{rhosa}). For the co-moving volume $V_h=\frac{4\pi}{3}a^3$, the above equation indicates $\rho_{p0}\,a^{-4}+\rho_s=0$ or $\rho_{p0}\,a^{-4}+\rho_s\sim a^{-3}$. Note that $\rho_{p0}\,a^{-4}+\rho_s=0$ means that the energy density of the substance is negative, which is non-physical. As for the latter case, $\rho_s$ will become negative as $a\rightarrow0$, which is also non-physical. Therefore, for the co-moving volume, the disappearance of the energy of the CMB photons cannot be solved by introducing the coupling between photons and other matter. For the apparent horizon ($V_h=\frac{4\pi}{3}H^{-3}$), according to Friedmann equation $3H^2=\kappa^2(\rho_{p0}\,a^{-4}+\rho_s+\rho_m)$ and $\rho_m\,V_h=\text{Const.}$, it can also be proved that Eq.~(\ref{eq210}) is invalid.

Since gravity is also evenly distributed inside the whole horizon, even if the energy density of the gravitational field can be well defined, according to the definition of the MS energy, it is clear that the energy density of the gravitational field of the system is also proportional to the radius of the horizon to the third power. In this way, one still gets Eq.~(\ref{eq210}) with $\rho_s$ being the energy density of the gravitational field. Therefore, it seems that the definition of the energy density of the gravitational field does not help much to solve the energy loss of the CMB photons.

Reviewing Friedmann equations, it is found that as long as $V_h\sim H^{-2}$, the total energy inside the horizon will be always a conserved quantity. Therefore, choosing a special horizon ($V_h\sim H^{-2}$) as the volume of the universe may solve the energy loss of the CMB photons. However, such a special horizon has two fatal problems. First, the number of particles inside the horizon is not conserved. To achieve energy conservation inside the horizon, one has to request that there exists energy (particles) exchange between the outside and inside of the horizon. Second, we now cannot figure out what its actual physical meaning is.

In summary, by investigating the red-shift of the CMB photons in the universe, it is found that from a phenomenological point of view, energy conservation cannot be achieved through the non-minimal coupling between matter fields or the definition of the energy density of the gravitational field. So what else can we do to solve this problem? Note that the above hypothetical unknown substance always spreads all over the universe, so Eq.~(\ref{eq210}) inevitably needs to be satisfied. But if the substance interacting with photons does not spread the entire universe and even it is not a three-dimensional object, what will happen? In this case, it is easy to associate the unknown substance with the two-dimensional horizon. If there exists a kind of energy related to the area of the horizon, we may solve the problem of energy conservation in the universe. On the other hand, we know that for an expanding system in classical thermodynamics, the energy loss of the system is due to the work done by the system on its surroundings, which is related to the surface area and the pressure of the system. For the co-moving volume in cosmology, if one counts the work done by photons (if it exists) as $W=p_p\,\text{d}V$ ($p_p$ is the pressure of photons), then the red-shift of photons is naturally caused by the work, and so there is no energy non-conservation in the universe. When energy conservation could be implemented in cosmology, it will be consistent with energy-momentum tensor conservation~\cite{Prigoginex1,Prigoginex2,Harko:2014pqa,Harko:2015pma}. Next, we will discuss related issues in detail, and further extend these studies to any horizons and the non-minimal coupling gravity.

\section{The correspondence between energy conservation and energy-momentum tensor conservation with surface energy and effective pressure}
\label{sec3}

In this section, we start from the volume enclosed by any cosmological horizons in the universe. By calculating the energy of the matter inside the horizon, we can define a surface energy on the horizon to implement energy conservation of the system (the energy inside the volume plus the energy on the horizon). In this case, the matter inside the horizon could be regarded as a freely expanding object and the surface energy on the horizon is the intrinsic energy of the horizon. If the matter inside the horizon could do work on its surroundings (i.e., it is not free to expand), then we can define an effective pressure at the horizon, which could also implement energy conservation of the system. Both definitions could ensure the correspondence between energy conservation and energy-momentum tensor conservation. We first focus on two types of special horizons: the apparent horizon and the co-moving horizon. At the end of the section, we generalize our discussions to an arbitrary horizon.

For the matter contained in any horizons, during the evolution of the scale factor from $a_1$ to $a_2$, the corresponding change in the MS energy is given as
\begin{eqnarray}
\nabla\mathfrak{E}_V=\rho_t(a_2)V_h(a_2)-\rho_t(a_1)V_h(a_1),\label{aaa}
\end{eqnarray}
where $\rho_t(a)=\rho_r(a)+\rho_d(a)+\rho_{DM}(a)+\rho_{DE}(a)$.

We define the surface energy on the horizon as a function of the scale factor $\mathfrak{E}_H(a)$. Then the change in the horizon energy from $a_1$ to $a_2$ can be given as
\begin{eqnarray}
\nabla\mathfrak{E}_H=\mathfrak{E}_H(a_2)-\mathfrak{E}_H(a_1).\label{aaa1}
\end{eqnarray}
For a given horizon, if $\nabla\mathfrak{E}_V=-\nabla\mathfrak{E}_H$ is fixed for any $a_1$ and $a_2$, we can state that the total energy of the gravitational system is conserved. Therefore, the definition of the horizon energy $\mathfrak{E}_H$ is totally dependent on the volume $V_h$ and all substances inside the horizon.

Similarly, in order to guarantee energy conservation (for the system and its outside), the definition of the effective pressure $\hat P$ at the horizon needs to satisfy
\begin{eqnarray}
\nabla\mathfrak{E}_V=-\int_{r_h(a_1)}^{r_h(a_2)} \hat P\,\text{d}V_h,\label{aaa3}
\end{eqnarray}
where $r_h(a)$ is the radius of the horizon.

\subsection{Apparent horizon}
\label{sec31}
The apparent horizon is the most researched horizon in cosmological thermodynamics. Since we only consider the universe with vanishing curvature, the radius of the apparent horizon is given as $r_h(a)=H^{-1}$, and the volume inside the apparent horizon is $V_h=\frac{4\pi}{3}H^{-3}$. With Friedmann equations and $V_h=\frac{4\pi}{3}H^{-3}$, Eq.~(\ref{aaa}) can be rewritten as
\begin{eqnarray}
\nabla\mathfrak{E}_V&=&\frac{4\pi}{3}\rho_t(a_2)H^{-3}(a_2)
-\frac{4\pi}{3}\rho_t(a_1)H^{-3}(a_1)\nonumber\\
&=&\frac{4\pi}{\kappa^2}\left[H^{-1}(a_2)-H^{-1}(a_1)\right]\nonumber\\
&=&\frac{4\pi}{\kappa^2}\left[r_h(a_2)-r_h(a_1)\right].\label{aaaa2}
\end{eqnarray}
Since we expect $\nabla\mathfrak{E}_V=-\nabla\mathfrak{E}_H$ for any $a_1$ and $a_2$, then we have
\begin{eqnarray}
\mathfrak{E}_H(a)=E_0-\frac{4\pi}{\kappa^2}r_h,\label{aaaa21}
\end{eqnarray}
which is the only proper definition for the surface energy of the apparent horizon. Note that $E_0$ is an unknown constant with a dimension of energy and it must ensure that $\mathfrak{E}_H(a)$ is positive. Such a definition is just for the establishment of mathematical equations, and so it lacks a physical basis without doubt. As a two-dimensional object, the apparent horizon is actually unconvincing to say that it has energy. In this work, we put aside the physical nature of the surface energy on the horizon. Therefore, we reiterate that the assumption here is only from a phenomenological consideration. In this case, the energy density of the apparent horizon is
\begin{eqnarray}
\rho_H(a)=\frac{E_0}{4\pi\, r_h^2}-\frac{1}{\kappa^2\, r_h},\label{aaaa22222221}
\end{eqnarray}
which only evolves with the radius of the apparent horizon.

With the definition of the surface energy on the apparent horizon, the energy conservation of the entire system can be expressed as
\begin{eqnarray}
\frac{4\pi}{3}\rho_t(a)\,r_h^{3}+E_0-\frac{4\pi}{\kappa^2}\,r_h=\text{Const.}\,.\label{aaaa2111}
\end{eqnarray}
Taking the derivative of the above equation with respect to $t$, with the help of Friedmann equations (\ref{eq:FriedmannEq}) and (\ref{eq:PressureEq}), one can naturally get
\begin{eqnarray}
\dot\rho_t(a)+3H[\rho_t(a)+p_t(a)]=0,\label{aaaa2112}
\end{eqnarray}
which is exactly the energy-momentum tensor conservation equation. It is actually a very obvious result, because Eq.~(\ref{aaaa2111}) itself is an identity. The derivation process is equivalent to using Friedmann equations (\ref{eq:FriedmannEq}) and (\ref{eq:PressureEq}) to derive the energy-momentum tensor conservation equation, which is an inevitable result. Therefore, the energy conservation mentioned above is always consistent with energy-momentum tensor conservation.

Next, we try to analyze and calculate the energy change of the universe (space-time) within the apparent horizon from the perspective of work. In this case, the universe does not expand freely, which is similar to the practice used in Refs.~\cite{Prigoginex1,Prigoginex2}, except that the co-moving volume is replaced by the apparent horizon. We employ $\hat P$ to represent the effective pressure at the apparent horizon, which may be related to the internal matter, but it does not have to be equal to the total pressure of the internal matter. Since the apparent horizon is increasing during the expansion of the universe, the matter inside the apparent horizon is doing work on the outside. From $a_1$ to $a_2$, the total amount of the work can be given as
\begin{eqnarray}
\nabla W_h=\int_{r_h(a_1)}^{r_h(a_2)} \hat P\,\text{d}V_h=\int_{r_h(a_1)}^{r_h(a_2)}4\pi\, r_h^2\,\hat P\,\text{d}r_h.\label{aaaa3}
\end{eqnarray}
The requirement $\nabla\mathfrak{E}_V=-\nabla W_h$ leads to
\begin{eqnarray}
\hat P=-\frac{1}{\kappa^2\,r_h^2}=-\frac{H^2}{\kappa^2}=-\frac{1}{3}\rho_t(a).\label{aaaa3334}
\end{eqnarray}
Such a result is a little bit beyond our expectations, because $\hat P$ is only a simple function of the Hubble rate. Since $\hat P$ is negative, the effective pressure is toward inside the apparent horizon. If the total pressure of all matter in the universe satisfies $p_{t}(a)=\omega(a)\rho_{t}(a)$ with $\omega(a)$ being a function of the scale factor, the relationship between the effective pressure $\hat P$ and the total pressure of all matter is given as
\begin{eqnarray}
\hat P=-\frac{1}{3}\rho_{t}(a)=-\frac{1}{3}\frac{p_{t}(a)}{\,\omega(a)}.\label{aaaa4}
\end{eqnarray}
This result manifests that when the matter inside the apparent horizon is only radiation, i.e., $\omega(a)=1/3$, the effective pressure at the apparent horizon $\hat P$ is equal to the pressure of radiation in the opposite direction.

When the effective pressure at the apparent horizon satisfies Eq.~(\ref{aaaa3334}), the thermodynamic energy equation $\text{d}E=-\hat P\,\text{d}V$ can be expressed as
\begin{eqnarray}
\text{d}E=\frac{1}{3}\rho_t(a)\,\text{d}V_h.
\end{eqnarray}
Since we require $\text{d}E$ to be equal to the energy loss of the system, then we have
\begin{eqnarray}
-\frac{1}{3}\rho_t(a)\,{\text{d}}V_h={\text{d}}[\rho_t(a)V_h].
\end{eqnarray}
Combined with Friedmann equations (\ref{eq:FriedmannEq}) and (\ref{eq:PressureEq}), it can be proved that the above equation is also the energy-momentum tensor conservation equation (\ref{aaaa2112}).

We briefly summarize the above results. If the surface energy on the apparent horizon is determined by Eq.~(\ref{aaaa21}), during the free expansion of the universe, the energy increase of the apparent horizon is equivalent to the energy lost by the matter inside the apparent horizon. Therefore, the total energy of the system remains conserved. In this case, the total energy of the matter insider the apparent horizon (and also the surface energy on the apparent horizon) is not proportional to the area of the apparent horizon, but the radius of the apparent horizon. Moreover, we can define an effective pressure at the apparent horizon as Eq.~(\ref{aaaa3334}), so the work done by the matter inside the horizon is given as Eq.~(\ref{aaaa3}), which ensures that the work is equal to the energy loss of the matter. Note that the work is a quantity related to the thermodynamic process and the equation of state of the internal matter, but the horizon energy is a thermodynamic quantity of state, which is only related to the radius of the apparent horizon. With these two definitions, for the system composed of the apparent horizon and its internal matter, energy conservation and energy-momentum tensor conservation are consistent.

\subsection{Co-moving horizon}
\label{sec32}
In this section, we study another special horizon, i.e., the co-moving horizon, whose radius is given as $r_h(a)=C_0\,a$ with a constant $C_0$. The co-moving volume is the most common thermodynamic system studied in cosmology in previous studies. The remarkable feature of the co-moving horizon is that the number of particles inside the co-moving horizon is conserved, provided there is no particle production and annihilation. In this work, we assume that the co-moving horizon also possesses energy and thermodynamic characteristics similar to the apparent horizon. The change in the MS energy inside the co-moving horizon, from $a_1$ to $a_2$, can be written as
\begin{eqnarray}
\nabla\mathfrak{E}_V=\frac{4\pi}{3}C_0^3\left[\rho_t(a_2)
a_2^3-\rho_t(a_1)a_1^3\right].\label{aaa44}
\end{eqnarray}
Similar to the previous approach, we first define a surface energy on the co-moving horizon as $\mathfrak{E}_H(a)$. When the scale factor evolves from $a_1$ to $a_2$, the energy change of the co-moving horizon should be equal to Eq.~(\ref{aaa44}). Therefore, by defining
\begin{eqnarray}
\mathfrak{E}_H(a)=E_0-\frac{4\pi}{3}C_0^3\,\rho_t(a)a^3=E_0-4\pi \frac{H^2}{\kappa^2}r_h^3,
\end{eqnarray}
one can easily get $\nabla\mathfrak{E}_H=-\nabla\mathfrak{E}_V$. Note that $E_0$ must ensure that $\mathfrak{E}_H(a)$ is positive. The energy density of the horizon is
\begin{eqnarray}
\rho_H(a)=\frac{E_0}{4\pi \,r_h^2}-\frac{H^2}{\kappa^2}r_h,\label{aaaa223441}
\end{eqnarray}
which is dependent on the Hubble rate and the radius of the co-moving horizon.

On the other hand, from the perspective of work, the effective pressure $\hat P$ at the co-moving horizon is determined by the following equation:
\begin{eqnarray}
\nabla\mathfrak{E}_V&=&\frac{4\pi}{3}C_0^3\big\{[\rho_r(a_2)+
\rho_d(a_2)+\rho_{DM}(a_2)+\rho_{DE}(a_2)]a_2^3\nonumber\\
&&-[\rho_r(a_1)+
\rho_d(a_1)+\rho_{DM}(a_1)+\rho_{DE}(a_1)]a_1^3\big\}\nonumber\\
&=&\frac{4\pi}{3}C_0^3\big\{[\rho_r(a_2)+\rho_{DE}(a_2)]a_2^3\nonumber\\
&&-[\rho_r(a_1)+\rho_{DE}(a_1)]a_1^3\big\}\nonumber\\
&=&-\int_{r_c(a_1)}^{r_c(a_2)}4\pi\, r_c^2\,\hat P\,{\text{d}}r_c,\label{aaa5555}
\end{eqnarray}
where the second equality is based on $\rho_d(a)+\rho_{DM}(a)\sim a^{-3}$. If $\hat P=\hat P_r+\hat P_{DE}$, it can be obtained quickly that $\hat P_r=p_r=\frac{1}{3}\rho_r$ and $\hat P_{DE}=p_{DE}=-\rho_{DE}$ (for $\rho_{DE}=\text{Const.}$). It is not difficult to prove that for most matter\footnote{It needs to satisfy the conservation equation $\dot \rho+3H(p+\rho)=0$, and the equation of state is given as $p=\omega_0\rho$, where $\omega_0$ is a constant. Then the energy density can be obtained as $\rho=\rho_0\,a^{-3(\omega_0+1)}$.} in the co-moving horizon,
\begin{eqnarray}
\hat P=p=\omega\, \rho
\end{eqnarray}
is the solution to Eq.~(\ref{aaa5555}).

Comparing Eqs.~(\ref{aaa5555}) and (\ref{aaaa4}), it is found that for the apparent horizon, the relationship between $\hat P$ and $p$ relies on the equation of state $\omega(t)$, and $\omega(t)$ could evolve over time. For the co-moving horizon, when $\omega(t)$ is a constant, the relationship between $\hat P$ and $p$ is irrelevant to $\omega(t)$. But if $\omega(t)$ evolves over time, to figure out $\hat P$, we have to get the energy density of the matter with respect to the scale factor according to the specific form of $\omega(t)$. Note that for the apparent horizon, when $\omega(t)<0$, $\hat P$ and $p$ have the same sign, and when $\omega(t)>0$, their signs are opposite. The physical explanation of such a result is as follows. The energy within the apparent horizon actually increases with the radius of the apparent horizon [see Eq.~(\ref{aaaa2})]. According to the thermodynamic energy equation, the surroundings of the apparent horizon should do work on the matter inside the horizon, so the effective pressure at the horizon needs to satisfy $\hat P<0$ even though the universe is expanding. Since the energy density of all matter is always positive, $\hat P=-\frac{1}{3}\rho_t(a)$ provides a guarantee for the negative value of $\hat P$. If the universe collapses, the direction of $\hat P$ is reversed. However, for the co-moving volume, when $\omega(t)$ is a constant, $\hat P$ and $p$ always have the same sign. And since the energy density is always positive, $\hat P$ is always positive when $\omega(t)>0$, which ensures that when the energy within the co-moving horizon is reduced, the matter inside the horizon will do work on its surroundings, and vice versa. One can refer to Refs.~\cite{Tolman11,Prigoginex1,Prigoginex2,Harko:2008qz,Bertolami:2008zh,Harko:2011kv,Harko:2014pqa,Harko:2015pma} for more discussions on the thermodynamic properties of the co-moving volume. The proof that energy conservation is consistent with energy-momentum tensor conservation is straightforward, which has also been confirmed in Refs.~\cite{Tolman11,Prigoginex1,Prigoginex2,Harko:2008qz,Bertolami:2008zh,Harko:2011kv,Harko:2014pqa,Harko:2015pma}, and so we will not repeat them here again.

\subsection{Arbitrary horizon}
\label{sec33}
Now, we consider an arbitrary cosmological horizon. Suppose the radius of the horizon $R(a)$ is a positive function of the scale factor. During the time period that the scale factor evolves from $a_1$ to $a_2$, the change in the MS energy inside the horizon is also given by Eq.~(\ref{aaa}) with $V_h=\frac{4\pi}{3}R^3$. Similarly, as long as we define the surface energy on the horizon as
\begin{eqnarray}
\mathfrak{E}_H(a)=E_0-\frac{4\pi}{3}\rho_t(a)R^3,
\end{eqnarray}
energy conservation is always satisfied. The second term on the right side of the definition could directly offset the energy of the internal matter, so the total energy of the system is $E_0$. The energy density of the horizon is given as
\begin{eqnarray}\label{iopuoi}
\rho_H(a)=\frac{E_0}{4\pi\, R^2}-\frac{1}{3}\rho_t(a)R.
\end{eqnarray}
Therefore, energy conservation equation is trivial: $E_0=\text{Const.}$, which is still consistent with the energy-momentum tensor conservation equation.

Let us analyze how to achieve the energy conservation of the system from the perspective of work. We still assume that the effective pressure at the horizon could be divided into several parts according to the internal composition. Therefore, the work done by the horizon, from $a_1$ to $a_2$, is given as
\begin{eqnarray}
\nabla W_h=\int_{R(a_1)}^{R(a_2)}4\pi\, R^2\,(\hat P_d+\hat P_r+\hat P_{DM}+\hat P_{DE})\,\text{d}R.\label{ccc1}
\end{eqnarray}
For the standard $\Lambda$CDM model, since the equations of state of all matter are constants, their energy densities have a similar form: $\rho\sim a^{-3(\omega+1)}$. We find that if $R(a)$ happens to be a polynomial of the scale factor: $R(a)=Y a^n$, where $Y$ is a positive constant and $n$ could be any constants, the requirement $\nabla W_h=-\nabla\mathfrak{E}_V$ will lead to a simple correlation between the effective pressure at the horizon and the pressure of the matter inside the horizon. Taking $V_h=\frac{4\pi}{3}R^3$ and $R(a)=Y a^n$ into $\nabla W_h=-\nabla\mathfrak{E}_V$, the effective pressure is given as
\begin{eqnarray}
\hat P\!&=&\!\hat P_d+\hat P_r+\hat P_{DM}+\hat P_{DE}\nonumber\\
\!&=&\!\sum_{i=d,r,DM,DE}\frac{-n+1+\omega_i}{n\,\omega_i}p_i
\,\,\,(n\neq0)\nonumber\\
\!&=&\!\frac{1-n}{n} (\rho_d+\rho_{DM})
-\frac{3n-4}{n}p_r+p_{DE}
\,\,\,(n\neq0).\label{ccc2}
\end{eqnarray}
When $n=0$, we have $\nabla W_h=-\nabla\mathfrak{E}_V=0$ directly. For $n=1$, the above formula reverts to the case of the co-moving horizon, i.e., $\hat P_d=\hat P_{DM}=0$, $\hat P_r=p_r$, and $\hat P_{DE}=p_{DE}$.

For a general radius $R(a)$, the requirement $\nabla W_h=-\nabla\mathfrak{E}_V$ can be expressed as
\begin{eqnarray}
&&\frac{4\pi}{3}\left[\rho_{t}(a_2)R^3(a_2)-\rho_{t}(a_1)R^3(a_1)\right]=\nonumber\\
&&-4\pi\int_{a_1}^{a_2}\hat P\, R^2\, R'\,\text{d} a,\label{ccc3}
\end{eqnarray}
where $\rho_{t}(a)=\rho_d(a)+\rho_r(a)+\rho_{DM}(a)+\rho_{DE}(a)$ and the prime denotes the derivative with respect to the scale factor. Then the universal relationship between $\hat P$ and other variables is
\begin{eqnarray}
\hat P&=&-\frac{1}{3}\frac{\text{d}\rho_t(a)}{\text{d}a} \frac{\text{d}a}{\text{d}R}R-\rho_t(a)\nonumber\\
&=&H[p_t(a)+\rho_t(a)]\frac{\text{d}t}{\text{d}R}R-\rho_t(a),\label{ccc4}
\end{eqnarray}
where the second equal sign is based on Friedmann equations (\ref{eq:FriedmannEq}) and (\ref{eq:PressureEq}). When $R(a)=Y a^n$, it reverts to Eq.~(\ref{ccc2}). It is found that if $\rho_t(a)\sim a^{-3(\omega_t+1)}$ and $\omega_t$ is a constant, there is an analytical solution of $\hat P=-p_t$, which requires the radius of the horizon satisfies
\begin{eqnarray}
R(a)\sim a^{\frac{1+\omega_t}{1-\omega_t}}.\label{ccc5}
\end{eqnarray}
If $\hat P=p_t$ and $\rho_t(a)\sim a^{-3(\omega_t+1)}$ with $\omega_t$ being a constant, the radius of the horizon degrades into the situation of the co-moving horizon:
\begin{eqnarray}
R(a)\sim a.\label{ccc6}
\end{eqnarray}

With Eq.~(\ref{ccc4}), the thermodynamic energy equation is
\begin{eqnarray}
\text{d}E=&&\!\!-\left\{H[p_t(a)+\rho_t(a)]\frac{\text{d}t}{\text{d}R}R-\rho_t(a)\right\}\text{d}V_h\nonumber\\
=&&\!\!\text{d}[\rho_t(a)V_h],
\end{eqnarray}
where $V_h=\frac{4\pi}{3}R^3$. By setting $\text{d}E=-\nabla W_h=\nabla\mathfrak{E}_V$, obviously one can still get
\begin{eqnarray}
\dot\rho_t(a)+3H[\rho_t(a)+p(a)]=0,\label{ccc7}
\end{eqnarray}
which corresponds to energy-momentum tensor conservation.

\section{The generalized second law of thermodynamics}
\label{sec4}
In previous research, we have discussed some special horizons, which are endowed with a surface energy to maintain energy conservation in the context of cosmology. In this section, we analyze whether these horizons are reasonable and appropriate from the perspective of thermodynamic entropy. We assume that there exists a kind of surface entropy matching the horizon, which is similar to the surface entropy of the apparent horizon~\cite{Bousso:2002ju}. Moreover, such surface entropy is compatible with the particle entropy inside the horizon, that is, they can be directly added as an extension quantity. In this way, by inspecting the evolution of the total entropy (the horizon entropy and the internal particle entropy), we can analyze, with the generalized second law of thermodynamics~\cite{Bekenstein:1974ax,Davies:1987ti,Wu:2007se,Wu:2008ir}, the rationality of the horizons mentioned above.

We denote the area entropy of the horizon as $S_A$, which can be similarly defined as the entropy of the cosmological event horizon~\cite{Bekenstein:1972tm,Bekenstein:1973ur,Bekenstein:1974ax,Davies:1987ti,Wu:2007se,Wu:2008ir}:
\begin{eqnarray}
S_A= \frac{k_B}{4\,l_p^2}A_h,\label{ddd1}
\end{eqnarray}
where $k_B$ is the Boltzmann constant, $A_h$ is the area of the horizon, and $l_p$ is the Planck length.

For the particle entropy inside the horizon, we only consider radiation, dust, and dark matter. The radiation entropy can be directly given by the entropy formula of photon gas. For dust and dark matter, one can assign a constant specific entropy (the entropy of a single particle) for them~\cite{Prigogine:1989zz,Calvao:1991wg}, and so the corresponding entropy change can be described by calculating the change in the total number of particles inside the horizon. As for the cosmological constant or dark energy, we can regard its entropy as zero.

As a result, the total entropy of the system can be expressed as
\begin{eqnarray}
\!\!\!S_t=&&\!\!S_A+S_{dDM}+S_r\nonumber\\
=&&\!\!4\pi\, r_h^2\cdot \frac{k_B}{4\,l_p^2}\!+\!\frac{4}{3}\pi\, r_h^3\cdot\sigma_x\, n_x\!+\!\frac{4}{3}\pi\,r_h^3\cdot\frac{4\pi\, k_B^2}{45c^3\,\hbar^3} T^3,\label{ddd2}
\end{eqnarray}
where $r_h$ is the radius of the horizon. Here, $n_x$ is the total number density of dust and dark matter, $\sigma_x$ is the weighted average of the specific entropy of dust and dark matter, and $T$ is the temperature of radiation. Since normal entropy is non-negative, $\sigma_x$ is a positive constant. In the standard $\Lambda$CDM model, we have $n_x\propto a^3=n_0\,a^3$ ($n_0$ is the current number density of dust and dark matter) and $T\propto a^{-1}=T_0\, a^{-1}$ ($T_0$ is the current temperature of radiation).

If we suppose that there is no entropy exchange between all matter and the horizon, then the generalized second law of thermodynamics would be that each component (dust and dark matter have been regarded as a whole) satisfies the second law of thermodynamics, respectively:
\begin{eqnarray}
S_A'=&&\!\!8\pi\,r_h\cdot \frac{k_B}{4\,l_p^2}\cdot\frac{\text{d}r_h}{\text{d}a}>0,\label{441}\\\label{88881}
S_{dDM}'=&&\!\!4\pi\,r_h^2\cdot\sigma_x\, n_0\,a^{-3}\cdot\frac{\text{d}r_h}{\text{d}a}\nonumber\\
&&\!\!-4\pi\, r_h^3\cdot\sigma_x \,n_0\, a^{-4}>0,\\\label{8888}
S_{r}'=&&\!\!4\pi\, r_h^2\cdot\frac{4\pi\,k_B^2}{45c^3\,\hbar^3} T_0^3\,a^{-3}\cdot\frac{\text{d}r_h}{\text{d}a}\nonumber\\
&&\!\!-4\pi\, r_h^3\cdot\frac{4\pi\, k_B^2}{45c^3\,\hbar^3} T_0^3\,a^{-4}>0,\label{ddd4}
\end{eqnarray}
where the prime still indicates the derivative with respect to the scale factor. Since $a(t)$ increases with $t$, $S'>0$ is equivalent to $\text{d} S/\text{d}t>0$. For an arbitrary $r_h$, the above conditions boil down to
\begin{eqnarray}
\frac{\text{d}r_h}{\text{d}a}>\frac{r_h}{a}>0,\label{ddd5}
\end{eqnarray}
which means that as long as the entropy of particles inside the horizon satisfies the second law of thermodynamics, the entropy of the whole system will be constantly increasing. Note that if the equation of state of dark energy is set as $\omega_{DE}<-1$, it will lead to $8\pi\, r_h\cdot \frac{k_B}{4\,l_p^2}\cdot\frac{\text{d}r_h}{\text{d}a}<0$ as $a\rightarrow\infty$. Therefore, we do not consider such a case. Moreover, since the above expression of entropy only corresponds to the late stage of the universe and the time after that, we can relax the second law of thermodynamics further as: $S'>0$ as $a\rightarrow\infty$.

When $r_h=H^{-1}$, it is easy to judge the authenticity of $\frac{\text{d}H^{-1}}{\text{d}a}>\frac{H^{-1}}{a}$. With the help of Eq.~(\ref{eq:FriedmannEq}), we just need to verify $\frac{\text{d}\rho_t^{-1/2}}{\text{d}a}>\frac{\rho_t^{-1/2}}{a}$. If $\rho_t\propto a^{-3(\omega_t+1)}$ with a constant $\omega_t$, then it requires $\omega_t>-\frac{1}{3}$. It is known that the $\Lambda$CDM model predicts that the universe will eventually be dominated by the cosmological constant, and so the equation of state of the universe will approach $-1$ infinitely as $a\rightarrow\infty$. When the universe evolves to the late stage, the entropy of particles inside the apparent horizon does not satisfy the second law of thermodynamics. Note that when the effective equation of state of the universe is larger than $-1$, $\frac{\text{d}H^{-1}}{\text{d}a}>0$ is always tenable. Therefore, the entropy of the apparent horizon is always increasing.

When $r_h\sim a$, the entropy of particles inside the horizon is a constant and the horizon entropy is increasing with the expansion of the universe. The generalized second law of thermodynamics will not be violated.

If these entropy can be converted to each other, then the generalized second law of thermodynamics requires
\begin{eqnarray}
S_t'=&&\!\!8\pi\, r_h\cdot \frac{k_B}{4\,l_p^2}\cdot\frac{\text{d}r_h}{\text{d}a}+\Big(4\pi\, r_h^2\cdot\sigma_x\, n_0\, a^{-3}\cdot\frac{\text{d}r_h}{\text{d}a}\nonumber\\
-&&\!\!4\pi\, r_h^3\cdot\sigma_x\, n_0\, a^{-4}\Big)
+\Big(4\pi\, r_h^2\cdot\frac{4\pi\, k_B^2}{45c^3\,\hbar^3} T_0^3\,a^{-3}\cdot\frac{\text{d}r_h}{\text{d}a}\nonumber\\
-&&\!\!4\pi\, r_h^3\cdot\frac{4\pi\, k_B^2}{45c^3\,\hbar^3} T_0^3\,a^{-4}\Big)>0.\label{ddd3}
\end{eqnarray}
It is worth noting that some studies showed that the entropy of dust and dark matter is much smaller than the entropy of radiation in a unit volume, so the particle entropy inside the horizon is mainly dominated by radiation (see Ref.~\cite{Egan:2009yy} and references therein).

When $r_h=H^{-1}$, we already know that the entropy of particles does not satisfy the second law of thermodynamics at the late stage of the universe (i.e., $a\rightarrow\infty$). Therefore, in order to ensure that the entire system meets the second law of thermodynamics, the entropy increase of the horizon must be greater than the entropy decrease of particles. According to Eqs.~(\ref{88881}) and (\ref{8888}), the first term in $S_{dDM}'$ ($S_r'$) is positive and the second term is negative. Therefore, as $a\rightarrow\infty$, if we have
\begin{eqnarray}
S_A'-4\pi\, r_h^3\cdot\sigma_x \,n_0\, a^{-4}-4\pi\, r_h^3\cdot\frac{4\pi\, k_B^2}{45c^3\,\hbar^3} T_0^3\,a^{-4}>0,
\end{eqnarray}
the entropy of the entire system absolutely satisfies the generalized second law of thermodynamics. Considering that the second terms in $S_{dDM}'$ and $S_r'$ have a same structure, we can define
\begin{eqnarray}
Q_0=\sigma_x \,n_0+\frac{4\pi\, k_B^2}{45c^3\,\hbar^3} T_0^3=\text{Const.}\,.
\end{eqnarray}
It can be proved that as long as
\begin{eqnarray}\label{fff880}
\frac{ \frac{k_B}{\,l_p^2}}{2 Q_0}\cdot\frac{1}{\sqrt{3}}\kappa\left(\frac{3\rho_{dDM0}}
{2\sqrt{\rho_{DE0}}}\right)>1,
\end{eqnarray}
we have
\begin{eqnarray}
S_A'-4\pi\, r_h^3\cdot Q_0\, a^{-4}>0.\label{ddd6}
\end{eqnarray}
Then, $S_t'>0$ can be satisfied at the late stage of the universe (see Appendix). As $a\rightarrow\infty$, the horizon entropy tends to a constant and the particle entropy tends to zero, so the dominant term of the total entropy of the universe is indeed the horizon entropy~\cite{Egan:2009yy}.

When $r_h\propto a$, the particle entropy inside the co-moving horizon is a constant, and so the entropy change is only dependant on the horizon entropy. It is obvious that Eq.~(\ref{441}) is always tenable as $r_h\propto a$. When $r_h$ is random, it is difficult to further simplify the constraint of the generalized second law of thermodynamics on the system [see Eq.~(\ref{ddd3})]. But as long as Eq.~(\ref{ddd5}) is satisfied, Eq.~(\ref{ddd3}) definitely holds.

\section{The correspondence in the non-minimal coupling gravity}
\label{sec5}
Based on previous research, it is found that we can select a specific horizon as the boundary of the universe and assume that the energy of the system includes the particle energy inside the horizon and the surface energy on the horizon. By defining the surface energy, the total energy conservation of the system can be accomplished. Although such a definition of the surface energy lacks a physical basis, the expansion of the universe can be regarded as free expansion while the energy loss of the universe can be explained. On the other hand, we find that for any horizons in the $\Lambda$CDM model, as long as the effective pressure $\hat P$ at the horizon meets Eq.~(\ref{ccc4}), the energy change of the matter inside the horizon is equal to the work done by the matter. In both cases, the energy conservation equation and the thermodynamic energy equation of the closed system are consistent with the energy-momentum tensor conservation equation, respectively. Next, we extend the relationship to the gravity theory with a non-minimal coupling between geometry and matter, in which energy-momentum tensor conservation is generally not valid.

For the non-minimal coupling gravity, the equation for the matter field can be rewritten as
\begin{eqnarray}
\dot \rho+3H(p+\rho)=\hat X,\label{ooo7}
\end{eqnarray}
where $\hat X$ indicates the interaction between geometry and matter.

This equation can correspond to the thermodynamic equation describing the energy change of an open system, which can be given as
\begin{eqnarray}
\text{d}(\rho/n)=\text{d}q-p\,\text{d}(1/n),\label{ooo8}
\end{eqnarray}
where $\text{d}q=\text{d}Q/N$ is the average heat received by a particle and $n=N/V$ is the particle number density of the matter. Therefore, $Q$ is the heat received by the system and $N$ is the total number of the system. When the system is the co-moving volume, one can interpret $\hat X$ as an extra pressure at the horizon or an extra energy density inside the horizon. If the pressure at the horizon is modified, there will be an extra pressure in Friedmann equations~\cite{Ford:1986sy,Traschen:1990sw,Abramo:1996ip,Zimdahl:1999tn}. And if the energy density changes while the pressure does not change, there will exist particle production~\cite{Prigoginex1,Prigoginex2,Harko:2014pqa,Harko:2015pma}. The two thermodynamic explanations of Eq.~(\ref{ooo7}) are actually cannot be distinguished from the macro level.

For a system with particle production but without heat exchange, we have $\text{d}q=\text{d}Q/N=0$, so Eq.~(\ref{ooo8}) can be reexpressed as
\begin{eqnarray}
\dot \rho-(p+\rho)\frac{\dot n}{n}=0,\label{ooo9}
\end{eqnarray}
where we have used Eq.~(\ref{ooo8}) divided by $\text{d}t$. By defining a particle production rate as
\begin{eqnarray}
\Gamma=\frac{\dot N}{N}=\frac{\dot n}{n}+\frac{\dot V}{V},\label{ooo10}
\end{eqnarray}
we have
\begin{eqnarray}
\dot \rho-(p+\rho)\left(\Gamma-\frac{\dot V}{V}\right)=0.\label{ooo11}
\end{eqnarray}
For the co-moving volume without particle production ($\Gamma=0$), it degrades into Eq.~(\ref{ccc7}). If there is particle production in the co-moving volume, we have
\begin{eqnarray}
\dot \rho+3H(p+\rho)=(p+\rho)\Gamma.\label{ooo12}
\end{eqnarray}
Comparing this equation with Eq.~(\ref{ooo7}), we have
\begin{eqnarray}
\hat X=(p+\rho)\Gamma,\label{ooo13}
\end{eqnarray}
which can be explained as the phenomenon of particle production (or annihilation) due to the non-minimal coupling between geometry and matter~\cite{Harko:2014pqa,Harko:2015pma}.

For the case of the co-moving volume without particle production ($\Gamma=0$) but with heat exchange, we can write Eq.~(\ref{ooo8}) as
\begin{eqnarray}
\dot \rho+3H(p+\rho)=n\frac{\text{d}q}{\text{d}t}.\label{ooo14}
\end{eqnarray}
Comparing the above formula with Eq.~(\ref{ooo7}), one can obtain
\begin{eqnarray}
\hat X=n\frac{\text{d}q}{\text{d}t}.\label{ooo15}
\end{eqnarray}
The physical meaning is that the interaction between geometry and the matter field causes the pressure of all particles to change, which is similar to a heat source. In most of the literature, the change in particle pressure caused by this kind of the interaction is also called the production pressure, which satisfies
\begin{eqnarray}
p_c=-\frac{n}{3H}\frac{\text{d}q}{\text{d}t}.\label{ooo16}
\end{eqnarray}
Therefore, Eq.~(\ref{ooo14}) can be expressed as
\begin{eqnarray}
\dot \rho+3H(p+p_c+\rho)=0.\label{ooo17}
\end{eqnarray}
Because both $n$ and $H$ are positive, when $\frac{\text{d}q}{\text{d}t}>0$, $p_c$ is negative, which can promote the expansion of the universe~\cite{Ford:1986sy,Traschen:1990sw,Abramo:1996ip,Zimdahl:1999tn}. Combining with Eqs.~(\ref{ooo13}), (\ref{ooo15}), and (\ref{ooo16}), the thermodynamic effect brought by $\hat X$ for different physical process (with or without particle production) satisfies the following relationship:
\begin{eqnarray}
p_c=-\frac{\Gamma(p+\rho)}{3H}=-\frac{\hat X}{3H}.\label{ooo18}
\end{eqnarray}

In summary, the discussions above are the correspondence between the classical thermodynamic formula of an open co-moving volume [see Eq.~(\ref{ooo8})] and the equation of motion of the matter [see Eq.~(\ref{ooo7})] in the context of the non-minimal coupling gravity. The interaction between geometry and matter can be explained as particle production (annihilation) induced by gravity [see Eq.~(\ref{ooo12})]. The particle production rate $\Gamma$ and the interaction quantity $\hat X$ satisfy Eq.~(\ref{ooo13}). If there is no particle production, one can also explain the interaction as exciting (restraining) the kinetic energy of particles by gravity. So the pressure of particles will be changed and the extra pressure is usually called the production pressure. The corresponding thermodynamic formula (\ref{ooo8}) can be written as Eq.~(\ref{ooo17}). The production pressure $p_c$ and the interaction quantity $\hat X$ satisfies Eq.~(\ref{ooo18}).

Next, we generalize these relationships from the co-moving volume to an arbitrary volume with a horizon radius $R(a)$. First of all, the equation of motion of the matter remains unchanged. For the horizon radius $R(a)$, Eqs.~(\ref{ooo12}) and (\ref{ooo14}) become
\begin{eqnarray}
\dot \rho+3\frac{\dot R}{R}(p+\rho)=(p+\rho)\Gamma=n\frac{\text{d}q}{\text{d}t}.\label{ooo19}
\end{eqnarray}
According to the idea that energy conservation and energy-momentum tensor conservation could be equivalent in standard cosmology, Eq.~(\ref{ooo19}) can also directly correspond to Eq.~(\ref{ooo7}), which yields
\begin{eqnarray}
\hat X&=&n\frac{\text{d}q}{\text{d}t}-3(p+\rho)\left(\frac{\dot R}{R}-H\right)\nonumber\\
&=&(p+\rho)\left(\Gamma-3\frac{\dot R}{R}+3H\right).\label{ooo20}
\end{eqnarray}

When the universe expands freely, we can also define a surface energy on the horizon to keep energy conservation in the non-minimal coupling gravity. For any horizons, energy conservation demands that the surface energy $\mathfrak{E}_H(a)$ satisfies
\begin{eqnarray}
\mathfrak{E}_H(a_2)-\mathfrak{E}_H(a_1)=-\frac{4\pi}{3}\rho(a_2)R^3(a_2)\nonumber\\
+\frac{4\pi}{3}\rho(a_1)R^3(a_1).\label{ooo23}
\end{eqnarray}
Therefore, we can define $\mathfrak{E}_H(a)$ as
\begin{eqnarray}
\mathfrak{E}_H(a)&=&E_0-\frac{4\pi}{3}\rho \,R^3\nonumber\\
&=&E_0-\frac{4\pi}{3}\left(\frac{\hat X-\dot \rho}{3H}-p\right)R^3,
\end{eqnarray}
where the second equality is based on Eq.~(\ref{ooo7}). The energy density of the horizon is given by
\begin{eqnarray}\label{ooo24}
\rho_H(a)=\frac{E_0}{4\pi\, R^2}-\frac{1}{3}\left(\frac{\hat X-\dot \rho}{3H}-p\right)R.
\end{eqnarray}
If the relationship between $\hat X$ and the scale factor is confirmed, we can obtain the energy density of the horizon with respect to the scale factor. In this case, energy conservation is trivial: $E_0=\text{Const.}$, which is obviously consistent with energy-momentum tensor conservation.

Moreover, from the perspective of work, the relationship between the effective pressure $\hat P$ at the horizon surface and the radius of the horizon should be similar to Eq.~(\ref{ccc3})\footnote{Note that the energy change of the system is always equal to the value of the work done by the matter, i.e., Eq.~(\ref{ccc3}), regardless of whether there exists a heat source or particle production.}, which is given as
\begin{eqnarray}
\hat P&=&-\frac{1}{3}\frac{\text{d}\rho}{\text{d}a} \frac{\text{d}a}{\text{d}R}R-\rho\nonumber\\
&=&-\frac{1}{3}\frac{R}{\dot R}\hat X+H\frac{R}{\dot R}(p+\rho)-\rho,\label{ccc2220004}
\end{eqnarray}
where $\hat X$ is related to the particle production rate or the heat source of the system given by Eq.~(\ref{ooo20}). With this definition, we can naturally get that the thermodynamic energy equation [see Eq.~(\ref{ccc3})] is consistent with the evolution equation of energy-momentum tensor [see Eq.~(\ref{ooo7})].

For the co-moving horizon, the effective pressure at the horizon surface is equal to the internal particle pressure plus the pressure related to $\hat X$, which has been verified in Refs.~\cite{Prigoginex1,Prigoginex2,Harko:2014pqa,Harko:2015pma}. For the apparent horizon, we have
\begin{eqnarray}
\hat P=\frac{1}{3}\frac{H}{\dot H}\hat X-\frac{H^2}{\dot H}(p+\rho)-\rho.\label{ooo22}
\end{eqnarray}
When $\hat X=0$, it degrades into Eq.~(\ref{aaaa3334}).

Finally, let us calculate the net energy change of the matter when there is only a heat source in the system. Note that the system includes the matter and the heat source, whose energy change is given by $\frac{4\pi}{3}\rho(a_2)R^3(a_2)-\frac{4\pi}{3}\rho(a_1)R^3(a_1)$. Therefore, the energy change of the system is equal to the work done by the effective pressure of the matter. Since there is a heat source, during the expansion of the system, the matter absorbs energy from the heat source while consuming energy through doing work. We define the net energy change of the matter as the energy absorbed from the heat source minus the work done by the matter.

Since $\text{d}q=\text{d}Q/N$ is the average heat received by a particle, the energy that the matter absorbs from the heat source can be given as
\begin{eqnarray}
\Delta Q\!&=&\!-\int_{t_1}^{t_2}\frac{\text{d}q}{\text{d}t}N \, \text{d}t\nonumber\\
\!&=&\!-\frac{4\pi}{3}\!\!\int_{R(a_1)}^{R(a_2)}\frac{R^3 }{\dot R}\!\left[\hat X\!+\!3(p\!+\!\rho)\!\left(\frac{\dot R}{R}\!-\!H\right)\!\right]\!\, \text{d}R.
\end{eqnarray}
Therefore, the net energy change of the matter is
\begin{eqnarray}
\Delta E&=&\Delta Q-\int_{R(a_1)}^{R(a_2)}\hat P\, \text{d}V\nonumber\\
&=&-\int_{R(a_1)}^{R(a_2)}\frac{4\pi}{3}\frac{R^3 }{\dot R}\left[\hat X+3(p+\rho)\left(\frac{\dot R}{R}-H\right)\right]\text{d}R\nonumber\\
& &-\int_{R(a_1)}^{R(a_2)}\left[-\frac{1}{3}\frac{R}{\dot R}\hat X+H\frac{R}{\dot R}(p+\rho)-\rho\right]\text{d}V\nonumber\\
&=&-\int_{R(a_1)}^{R(a_2)}4\pi\, p\, R^2\,\text{d}R\nonumber\\
&=&-\int_{R(a_1)}^{R(a_2)} p\,\text{d}V.\label{uuu}
\end{eqnarray}
The above formula shows that the net energy change of the matter (not the system) is equal to the work done by the internal pressure of the matter (not effective pressure). In GR, since there are no heat sources, if the universe is not expanding freely, the net energy change of the matter inside a cosmological horizon is equal to the work done by the effective pressure. If the cosmological horizon is not the co-moving horizon, the effective pressure is generally not equal to the internal pressure of the matter. In the non-minimal coupling gravity, if the universe is not expanding freely, the total energy change of the system (which includes the matter and heat source) is also equal to the work done by the effective pressure. But the net energy change of the matter (deducting the heat source supplement) is given as Eq.~(\ref{uuu}). Therefore, as long as the pressure of the matter inside the horizon is positive (which means the net energy change of the matter is negative), all the energy provided by the heat source to the matter would be used for doing work, while the matter itself also loses energy due to the expansion of the system.

\section{Discussions and conclusions}
\label{sec6}
In cosmology, we usually choose the co-moving volume as a thermodynamic system to study the thermodynamic properties of the universe. The advantages of the co-moving volume are twofold. On the one hand, it guarantees that the number of particles in the co-moving volume is conserved when there is no particle production and annihilation. On the other hand, the boundary pressure of the co-moving volume is equal to the pressure of the internal matter (which is in accordance with the laws of classical thermodynamics). The disadvantage of the co-moving volume is that the boundary of the co-moving volume lacks clear physical meaning and so we do not know exactly where the boundary of the co-moving volume is. Given that, we study the thermodynamic properties and conservativeness of the system covered by an arbitrary horizon and generalize these studies to the non-minimal coupling gravity.

From the CMB photons, we analyzed why it is unlikely to implement energy conservation in the universe by introducing the coupling between matter fields or defining the energy density of the gravitational field. It is found that if there exists a kind of energy related to the boundary area of the system, the energy of the system could be conserved while the expansion of the universe could be completely free. By defining the surface energy density of the cosmological horizon [see Eq.~(\ref{iopuoi})], the energy of the matter inside the cosmological horizon will not disappear or be generated out of air. Then the energy conservation equation of the system is consistent with  the energy-momentum tensor conservation equation of the matter. If we discard the idea that the universe is expanding freely, we can suppose an effective pressure at the cosmological horizon [see Eq.~(\ref{ccc4})] to guarantee that the energy loss of the system within the horizon is equal to the work done by the matter~\cite{Tolman11,Prigoginex1,Prigoginex2}. Then the thermodynamic energy equation of the system is still consistent with the energy-momentum tensor conservation equation of the matter.

We investigated whether the surface entropy of an arbitrary cosmological horizon and the particle entropy inside the horizon satisfy the generalized second law of thermodynamics. We found that, for the apparent horizon and the co-moving horizon, the conditions for the system satisfying the generalized second law of thermodynamics are closely related to the components of the universe and the evolution of the horizon radius. As for the general cosmological horizon, it is difficult to simplify Eq.~(\ref{ddd3}). But if Eq.~(\ref{ddd5}) is tenable, the system definitely satisfies the generalized second law of thermodynamics.

Finally, we studied the thermodynamic energy equation and the evolution equation of energy-momentum tensor for any cosmological horizons in the non-minimal coupling gravity. We can still define a surface energy density on the cosmological horizon [see Eq.~(\ref{ooo24})] or an effective pressure at the horizon [see Eq.~(\ref{ccc2220004})] to implement general energy conservation. Then the thermodynamic energy equation of the system has the characteristics of an open thermodynamic system, which can be consistent with the evolution equation of the energy-momentum tensor of the matter. We calculated the net energy change of the matter (not the system, which consists of the matter and the heat source). It is found that when the pressure of the matter is positive, all the energy that the matter obtains from the heat source is used for doing work, and the matter itself also consumes internal energy for doing work, which means that the matter cannot actually gain energy from the heat source.

The present work, from a phenomenological perspective, showed that (thermodynamic) energy conservation and energy-momentum tensor conservation always do not conflict even for the system covered by an arbitrary cosmological horizon. The key to the consistency is to implement the energy conservation of the system. However, due to the fact that the number of particles is not conserved for the system covered by the general horizon, it seems counterintuitive to regard the system covered by an arbitrary horizon as a closed thermodynamic system. Therefore, although the thermodynamic conservation equation of the system can correspond mathematically to the energy-momentum tensor conservation equation of the matter, it needs more physical explanation, which is the major drawback of this research and also is a question worthy of our follow-up study. For the non-minimal coupling gravity, there could exist particle production and annihilation, so it may alleviate the problem about the non-conservation of the number of particles for an arbitrary horizon. To calculate the particle entropy within the horizon in the non-minimal coupling gravity, one needs to make some specific assumptions~\cite{Harko:2008qz,Bertolami:2008zh,Harko:2011kv}. In the end, the surface energy may provide ideas for solving the energy conservation problem in GR, which is also a topic that we can study in the future.

\section*{Appendix}
\label{sec7}
\appendix
Since we have $S_A'=8\pi\, r_h\cdot \frac{k_B}{4\,l_p^2}\cdot\frac{\text{d}r_h}{\text{d}a}>0$ and $4\pi\, r_h^3\cdot Q_0\, a^{-4}>0$ for $\omega_{DE}\geq-1$, if we can prove that as $a\rightarrow\infty$,
\begin{eqnarray}\label{xxx}
\frac{8\pi\,r_h\cdot \frac{k_B}{4\,l_p^2}\cdot\frac{\text{d}r_h}{\text{d}a}}{4\pi\, r_h^3\cdot Q_0\, a^{-4}}=
\frac{ \frac{k_B}{\,l_p^2}\cdot\frac{\text{d}r_h}{\text{d}a}}{2 r_h^2\cdot Q_0 \,a^{-4}}>1,
\end{eqnarray}
then Eq.~(\ref{ddd6}) is true, which naturally outputs $S_t'>0$ as $a\rightarrow\infty$. Taking $r_h=H^{-1}=\sqrt{3}\kappa^{-1}[\rho_{d0} \,a^{-3}+\rho_{r0}\,a^{-4}+\rho_{DM0}\, a^{-3}+
 \rho_{DE0}\, a^{-3(\omega_{DE}+1)}]^{-\frac{1}{2}}$ into Eq.~(\ref{xxx}) yields
\begin{eqnarray}
&&\!\frac{ \frac{k_B}{\,l_p^2}}{2 Q_0}\cdot\frac{1}{\sqrt{3}}\kappa\cdot\nonumber\\
&&\!\left[\frac{3\rho_{dDM0}+4\rho_{r0}\, a^{-1}+3\hat\omega\,\rho_{DE0}\, a^{3-3\hat\omega}}{2\sqrt{\rho_{dDM0} \, a^{-3}+\rho_{r0}\, a^{-4}+
 \rho_{DE0}\, a^{-3\hat\omega}}}\right]>1,\label{fff1}
\end{eqnarray}
where $\hat\omega=\omega_{DE}+1$ and $\rho_{dDM0}$ is the total energy density of dust and (cold) dark matter at present. For the $\Lambda$CDM model with $\omega_{DE}=-1$, as $a\rightarrow\infty$, to satisfy the above inequality we need
\begin{eqnarray}\label{fff88}
\frac{ \frac{k_B}{\,l_p^2}}{2 Q_0}\cdot\frac{1}{\sqrt{3}}\kappa\left(\frac{3\rho_{dDM0}}{2\sqrt{\rho_{DE0}}}\right)>1.
\end{eqnarray}
For dark energy density satisfying $\omega_{DE}>-1$, it can be proved that the left side of the inequality~(\ref{fff1}) satisfies
\begin{eqnarray}
&&\lim_{a\rightarrow\infty}\frac{ \frac{k_B}{\,l_p^2}}{2 Q_0}\cdot\frac{1}{\sqrt{3}}\kappa\cdot\nonumber\\
&&\left[\frac{3\rho_{dDM0}+4\rho_{r0}\, a^{-1}+3\hat\omega\,\rho_{DE0}\, a^{3-3\hat\omega}}{2\sqrt{\rho_{dDM0} \, a^{-3}+\rho_{r0}\, a^{-4}+
 \rho_{DE0}\, a^{-3\hat\omega}}}\right]\nonumber\\
&& =\frac{ \frac{k_B}{\,l_p^2}}{2 Q_0}\cdot\frac{1}{\sqrt{3}}\kappa
\left[\frac{3\rho_{dDM0}+3\hat\omega\,\rho_{DE0}\, a^{3-3\hat\omega}}{2\sqrt{0_{+}}}\right]\nonumber\\
&& =\infty>0,\label{fff22}
\end{eqnarray}
where $0_{+}$ indicates a positive infinitely small quantity. Therefore, when $\omega_{DE}\geq-1$ and Eq.~(\ref{fff88}) holds, Eq.~(\ref{ddd6}) is always true.

\section*{Acknowledgments} \hspace{5mm}
This work was supported by the National Natural Science Foundation of China under Grant Nos.~11873001 and 12005174. J. Li acknowledges the support of Natural Science Foundation of Chongqing (Grant No. cstc2018jcy-jAX0767).

\providecommand{\href}[2]{#2}\begingroup\raggedright

\end{document}